\documentclass[11pt, a4paper]{scrartcl}

\usepackage{latexsym}
\usepackage{amssymb}
\usepackage{amsmath}
\usepackage[authoryear, round]{natbib}

\usepackage{color}
\definecolor{myred}{rgb}{0.5,0,0}
\definecolor{myblue}{rgb}{0,0,0.75}
\definecolor{mygreen}{rgb}{0,0.5,0}

\usepackage{ifpdf}

\ifpdf
   \usepackage[pdftex]{graphicx}
   \usepackage[pdftex,
               pdftitle={Bounds for rating override rates},
               pdfsubject={Credit rating, rating override, discriminatory power, accuracy ratio, misclassification rate},
               pdfkeywords={},
               pdfauthor={Dirk Tasche},
               pdfstartview=FitR,
               breaklinks=true,
               colorlinks=true,
               citecolor=myred,
               linkcolor=myblue,
               urlcolor=mygreen]{hyperref}
\else
   \usepackage[dvips]{graphicx}
   \usepackage{hyperref}
   \usepackage{rotating}
\fi

\newtheorem{theorem}{Theorem}[section]

\newtheorem{remark}[theorem]{Remark}
\newtheorem{example}[theorem]{Example}
\newtheorem{proposition}[theorem]{Proposition}

\newtheorem{corollary}[theorem]{Corollary}

\numberwithin{equation}{section}

\title{Bounds for rating override rates}

\author{%
Dirk Tasche\thanks{E-mail: dirk.tasche@gmx.net\newline
The author currently works at the UK Financial Services Authority. 
The opinions expressed in this paper are those of the author 
and do not necessarily reflect views of the Financial Services Authority.}}

\date{First version: March 10, 2012\\
	This version: July 13, 2012}

\begin{document}
\maketitle

\begin{abstract}
Overrides of credit ratings are important correctives of ratings that are determined by 
statistical rating models. Financial institutions and banking regulators agree on this 
because on the one hand errors with ratings of corporates or banks can have fatal consequences for the 
lending institutions and on the other hand errors by statistical methods can be minimised but 
not completely avoided. Nonetheless, rating overrides can be misused in order to conceal the real
riskiness of borrowers or even entire portfolios. That is why rating overrides usually
are strictly governed and carefully recorded. It is not clear, however, which frequency of overrides is 
appropriate for a given rating model within a predefined time period. This paper
argues that there is a natural error rate associated with a statistical rating model that
may be used to inform assessment of whether or not an observed override rate is adequate.
The natural error rate is closely related to the rating model's discriminatory power
and can readily be calculated.
\end{abstract}


\section{Introduction}
\label{tas_sec_0}

Overrides of credit ratings generated by statistical models are a somewhat controversial topic. 
Financial institutions and regulators alike acknowledge that statistical models while being useful
for the acceleration of rating processes and quantification of rating results, in principle are
still inferior to careful expert assessments of creditworthiness. Nonetheless, even experienced credit experts
could undeliberately be biased to under- or overestimate the creditworthiness of certain borrowers
or groups of borrowers.

The following comment by the Basel Committee on Banking Supervision \citep[][extract from paragraph~417]{BaselAccord}
confirms and extends these observations:\\
``Credit scoring models and other mechanical rating procedures generally use only a subset of
available information. Although mechanical rating procedures may sometimes avoid some of
the idiosyncratic errors made by rating systems in which human judgement plays a large
role, mechanical use of limited information also is a source of rating errors. Credit scoring
models and other mechanical procedures are permissible as the primary or partial basis of
rating assignments, and may play a role in the estimation of loss characteristics. Sufficient
human judgement and human oversight is necessary to ensure that all relevant and material
information, including that which is outside the scope of the model, is also taken into
consideration, and that the model is used appropriately.'' 

While it is generally expected that overrides improve the quality of rating assignments\footnote{%
The empirical evidence of the impact of overrides on rating performance is not
unambiguous. Thus, in a recent study \citet{brown2012information} found that ``overall, 
our results suggest that the widespread use of discretion by loan officers may not
result in more accurate assessments of the creditworthiness of borrowers''. This finding,
however, might be a consequence of the fact that most of the banks providing data for the
study were not subject to the minimum requirements of the Basel~II Internal Ratings Based Approach.}, 
the Basel Committee
has a clear view on the need to monitor overriding activity \citep[][paragraph~428]{BaselAccord}:\\
``For rating assignments based on expert judgement, banks must clearly articulate
the situations in which bank officers may override the outputs of the rating process, including
how and to what extent such overrides can be used and by whom. For model-based ratings,
the bank must have guidelines and processes for monitoring cases where human judgement
has overridden the model's rating, variables were excluded or inputs were altered. These
guidelines must include identifying personnel that are responsible for approving these
overrides. Banks must identify overrides and separately track their performance.''

This paper is about analysing the performance of the overrides of the output of statistical
rating models. How to do this is by no means obvious. Look for instance at the case where
a borrower's rating is suggested to be poor by a statistical model but is overridden to a 
high quality rating grade. Assume that the borrower afterwards defaults within the following
twelve months. At first glance, this incident might be considered clear evidence of an unjustified
override. It could, however, just be an occurrence of bad luck. Only if we observed a significant
number of such outcomes of overrides we would be able to draw a conclusion on systematic bias
of the overrides.

From this example, we can derive a first performance criterion for rating overrides: The discriminatory 
power of a rating system post overrides should not be lower than the discriminatory power
of the system without the overrides. Another criterion, also based on a significant sample
of overrides, relates to the tendency of the overrides: If the majority of the overrides is towards
rating grades indicating poorer credit quality one has to investigate whether or not this
is caused by underestimation of the probabilities of default (PDs)\footnote{%
In the following we will often use the acronym PD for probability of default.} associated with the grades of
the rating system. 
Careful analysis should be applied to the reasons quoted as causing the overrides:
They could give indications of risk drivers not captured by the statistical model or no longer
being predictive of creditworthiness. 

Financial institutions should also look at the frequency of overrides applied to the outcomes
of rating models. Intuitively, observing a high frequency of overrides might indicate that something
is wrong with the model. 
This is why most financial institutions monitor the override frequencies of their wholesale rating 
systems and, typically, trigger corrective actions when pre-defined bounds 
for the override rates are exceeded.

\begin{figure}[t!p]
\caption{Unconditional and conditional rating distributions. The distributions conditional
on default and survival respectively have been inferred from the unconditional distribution under
the assumption that the unconditional PD is 5\% and the accuracy ratio associated with the rating 
model is either low at 25\% or high at 75\%. See example~\ref{ex:discrete} for 
details of the calculations.}
\label{fig:1}
\begin{center}
\ifpdf
	\includegraphics[width=15cm]{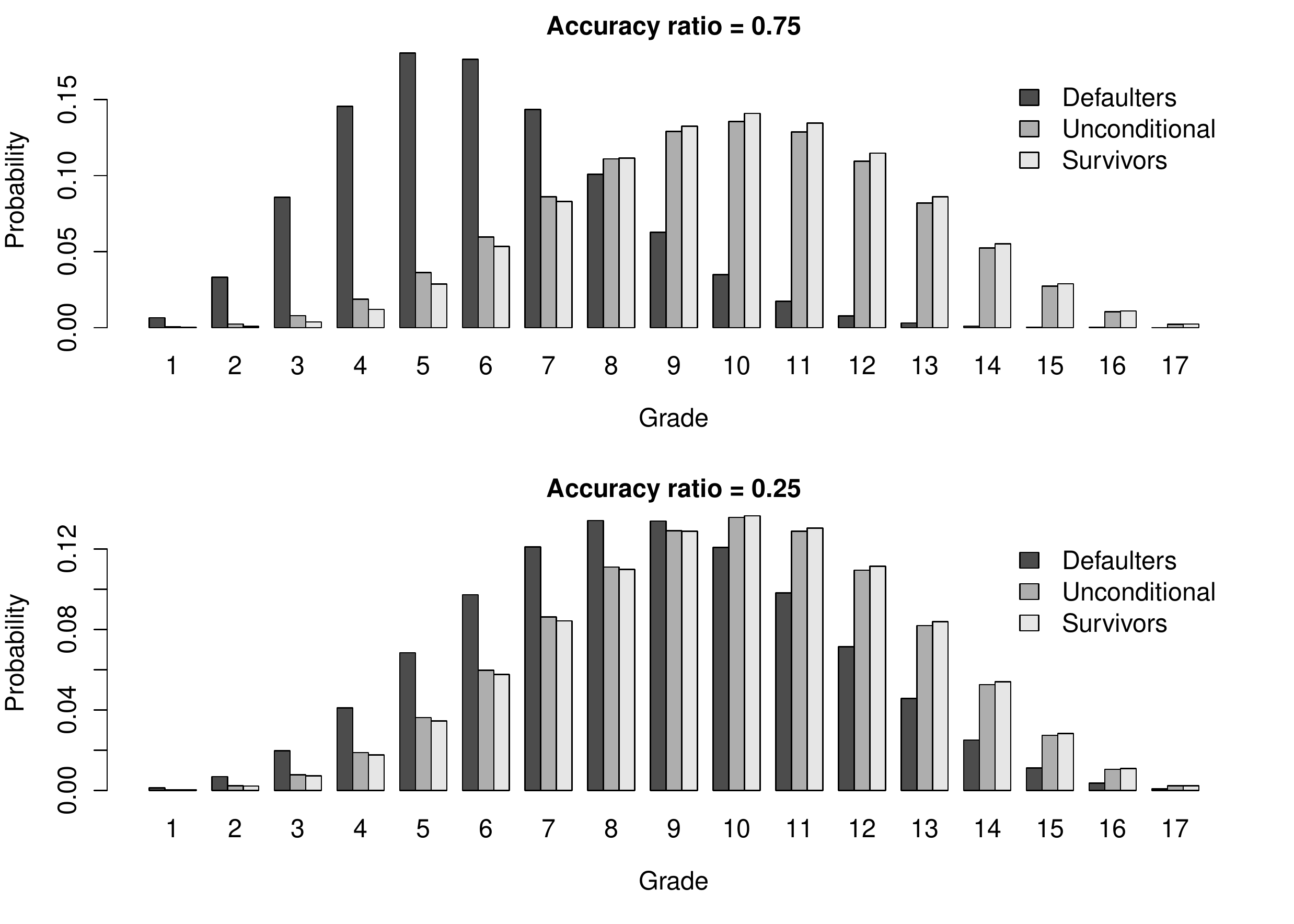}
\fi
\end{center}
\end{figure}

What, however, does ``high frequency'' mean? Would it be 10\% or rather 40\% of
the ratings assigned within a year?
By intuition one would say that the ``critical'' frequency of overrides to indicate
problems with a statistical rating model should be related to the discriminatory power of the model.
As demonstrated in figure~\ref{fig:1}, a model with low power needs more corrections 
-- in the shape of overrides -- than a model with 
high power because the overlap between the rating profiles of defaulting and solvent borrowers is
larger for low power rating models. In this paper, we suggest a method to determine bounds for override rates
that should not be exceeded if all were right with the underlying statistical model. 
We then investigate the link between the discriminatory power of the statistical rating model 
and the proposed rating override bounds, finding that indeed the bounds are the tighter
the more powerful the model is.

We also demonstrate how ex ante estimates of discriminatory power can be inferred from the
PD curves associated with statistical rating models. Such estimates may then be used to derive
the override rate bounds needed for effective monitoring. This is particularly useful 
in the case of low default portfolios where -- due to the lack of default observations -- 
monitoring and validation to a large extent must be based on the analysis of overrides. 

The paper is organised as follows: 
In section~\ref{tas_sec_1} we discuss the connection between rating error rates and 
the statistical notion of misclassification rate. We argue that the misclassification rate 
related to a particular classification rule provides 
a suitable approximate
upper bound for the rating error rate and therefore also for the rate of overrides of the
outputs of a statistical rating model. In section~\ref{tas_sec_2} we describe how to calculate the 
override rate bounds in the special cases of a rating model with a finite number of grades
and a rating model with normally distributed conditional score distributions. The
case of the conditional normal distributions, in particular, allows us to study the connection between override rate 
bounds and discriminatory power. We also present some numerical examples to
illustrate this connection. In section~\ref{tas_sec_monitoring} we discuss how
the proposed bound for rating override rates fits into an illustrative framework
for the monitoring of rating overrides. 
Section~\ref{tas_sec_4} summarises and concludes the paper.

\textbf{A note on the terminology.} In this paper, we study 
rating overrides in the context of rating systems with a small
finite number $k$ of grades\footnote{%
In order to achieve comparability of their internal ratings 
to agency ratings, financial institutions sometimes $k$ chose $k$ as seven, 
the number of performing unmodified grades used by the major rating agencies, or as 
seventeen, the number of performing modified grades used by the agencies. In the latter
case, grade 17 would correspond to S\&P grade AAA, grade 16 to S\&P grade AA$+$ etc.
until grade 1 that would correspond to S\&P grade CCC.}. 
For the purpose of the paper, it is assumed that a \emph{statistical model} is used to determine \emph{proposed
ratings} which may then be confirmed or overridden by experts to give the 
\emph{final ratings}. The direct output of the statistical model is called \emph{score} and may
be discrete with a large (compared to the number of rating grades $k$) range of scores or
on a continuous scale. 
The range of the scores is decomposed into a number of disjoint intervals each of which is then
mapped to one of the rating grades, thus generating the proposed ratings. 
The decomposition of the score range for the mapping to the 
grades usually is based on the probabilities of default associated with the scores.

\textbf{Convention.} Low values of the model output score indicate low creditworthiness, 
high values of the score indicate high creditworthiness. The mapping of the scores
to the rating grades (expressed as positive integers) is increasing, i.e.\ high grades
mean high creditworthiness.


\section{Approximating rating error rate by misclassification rate}
\label{tas_sec_1}

In the following, we assume that for every real-life rating model 
there are unavoidable rating errors because defaults are unpredictable.
We want to determine an approximation for the proportion of such
unavoidable rating errors compared to all rating actions. We also assume that
the rating models we consider are correctly calibrated such that
no overrides are needed in order to account for PD estimation bias.
For the purpose of this paper, such overrides would be considered 
avoidable (by recalibration of the rating model). Establishing
bounds for unavoidable errors will help to identify overrides 
that are due to miscalibration and thus in principle avoidable.

\subsection{The typical process for rating overrides}
\label{sec:mechanics}

Let us look at a -- typically wholesale (e.g.\ corporates or
financial institutions) -- portfolio of borrowers that are
in the scope of application of the rating system under consideration. A
rating action for one of the borrowers consists of three steps:
\begin{itemize}
	\item For the first step, a statistical model 
	is applied to data \emph{(risk factors)} related to the borrower, typically financial ratios from
	the borrower's balance sheet and/or qualitative assessments by credit officers
	marked up on a standard scale. The result of this step is a score $s$.
	\item For the second step, the score $s$ is mapped (e.g.\ by means
	of a look-up table) to a grade $g^\ast \in \{1, \ldots, k\}$. The grade $g^\ast$
	is called \emph{proposed rating}.
	\item In the third step, the rating proposal $g^\ast$ is reviewed by a credit expert
	or a committee of credit experts. The experts can decide to accept the rating
	proposal and assign the borrower the \emph{final rating} $g=g^\ast$. But 
	the experts can also decide to reject the rating proposal and assign the borrower a final
	rating $g \not= g^\ast$ (but with $g \in \{1, \ldots, k\}$).
	If a proposed rating is rejected, the experts have to record both the proposed 
	and the final ratings and the rationale for the rejection and the choice of the
	final rating.
\end{itemize}
The occurrence of $g \not= g^\ast$ in step three is called \emph{override} (of the proposed
rating). In most credit institutions overrides are subject to certain restrictions
which could include the following:
\begin{itemize}
	\item Overrides are not allowed if the proposed ratings are better than or equal to
	some threshold grade $k^\ast < k$. This rule might be established when it is felt that
	the relative differences between the better rating grades are so small that the grades
	cannot be meaningfully differentiated by the human mind.
	\item Overrides are only allowed if the final ratings are significantly different to
	the proposed ratings, e.g.\ if $|g - g^\ast| \ge b$ for some threshold value $b$. Such a rule
	might again be based on the intuition that the human mind cannot differentiate between
	adjacent rating grades.
	\item Only downgrade overrides are allowed. This rule would express a desire for rating
	conservatism.
\end{itemize}
If $n$ is the number of rating actions\footnote{%
The same borrower might be rated several times but usually each borrower needs
to be rated at least once per year.} with the rating system under consideration within
a pre-defined time period (e.g.\ one year), resulting
in pairs $(g^\ast_1, g_1), \ldots, (g^\ast_n, g_n)$ of proposed and final ratings,
then the \emph{override rate} is defined as the ratio of the number of rating actions
with different proposed and final ratings and the total number of rating actions:
\begin{equation}\label{eq:rate}
	\text{Override rate}\ = \ \frac{\#\{i: g^\ast_i \not= g_i\}}n.
\end{equation}

\subsection{`Right' and `wrong' ratings}
\label{eq:benchmark}

From \eqref{eq:rate} we get the obvious upper bound of 100\% for override rates -- not
very satisfactory. It might then seem a natural idea to try to derive an ``expected'' 
override rate for a statistical rating model by analysing the decision processes of the
credit experts charged with reviewing the rating proposals. One could argue that 
credit experts are likely to base their override decisons on comparisons to borrowers rated
in the past. 
This would mean that the experts look for similarities between the borrower to be reviewed and
the borrowers rated earlier. Such similarities might relate to financial conditions, management
quality or other descriptive information about the borrowers. 

Hence the procedure by credit experts to decide about possible overrides might be described in
mathematical terms as a methodology similar to the ``$m$ nearest neighbours'' ($m$-NN) methodology 
\citep[see, e.g.,][section~5.3]{Hand97}, which is well-known in fields like Pattern Recognition or 
Statistical Classification.
However, this analogy is not a promising path to the prediction of override rates. Assuming
we were able to identify the number $m$ of neighbours to be included in the $m$-NN analysis,
we would end up with developing another rating system especially for the override decision process.
But there is no reason to be sure that such a competitor rating would be more predictive of creditworthiness
than the statistical model under review. 

A more promising approach to bounds for override rates is to start from an interpretation of 
overrides as corrective actions. Then all 'wrong' -- in a sense to be determined -- 
ratings proposed by the statistical model should be overridden. Hence the 
proportion of wrong proposed ratings would be the `natural' override rate. As in practice
it is impossible to identify wrong ratings with certainty the natural override rate would
be an upper bound for the number of overrides.
This follows from the following considerations:
\begin{itemize}
	\item Assume the observed override rate exceeds the natural override rate. Then there are two
	possible explanations for this. The credit experts could have overridden too many ratings and,
	hence, have turned some correct ratings into wrong ratings. Or the proportion of wrong ratings
	is higher than expected and, hence, the discriminatory power of the statistical model is lower
	than predicted. Both explanations imply a deficiency in the rating process that should be rectified.
	\item Assume the observed override rate is less than the natural overide rate. Again there
	are two possible explanations. The credit experts could have overridden too few ratings and, hence,
	have failed to amend all wrong ratings. Or the proportion of wrong ratings is lower than expected and,
	hence, the discriminatory power of the statistical model is higher than predicted. While the first
	explanation may or may not indicate a deficiency in the rating process, the second explanation actually would
	be good news. 
	\item With a view on the potential consequences of these two cases (``override rate exceeds 
	the natural override rate'' and ``override rate is below the natural override rate''), we
	observe that the first case is clearly more of a concern than the second case. For the first case 
	indicates with certainty a deficiency of the rating process while in the second case there is a possibility
	that the statistical model actually is better than expected. This is why in practice the natural override
	rate can only be regarded as an upper bound for the observed override rate whose breach should trigger
	corrective action. 
\end{itemize}

So far, so well -- but how do we know which proportion of rating proposals is wrong?
Clearly, if we knew which would be the right -- or maybe only the most predictive -- ratings
we would be done because then we could insert the right ratings as the final ratings $g_i$ in 
\eqref{eq:rate}. Interestingly enough, if we assume that the most predictive ratings can
be determined based on the analysis of a finite number of risk factors, then in theory a
most predictive statistical rating model exists.
This follows from the Neyman-Pearson lemma \citep[see, e.g.,][Theorem 8.3.12]{Casella&Berger} that identifies
the most predictive statistic for the test of two simple hypotheses as the ratio of 
the two multi-variate joint densities of the risk factors on the defaulter population and
the survivor population respectively. Unfortunately, even if it all relevant risk factors 
could be identified, in practice it would be impossible to estimate accurately enough the two risk factor
densities.

Alternative candidates for the most predictive ratings might be agency ratings as well as 
the results of expert ranking exercises.
However, both agency ratings and expert rankings are not always available. More importantly, 
even if the quality of agency ratings for corporates is good in general, 
there is no reason to consider them the `right' ratings, implying the `right' ranking. 
If for nothing else, this follows from the fact that ratings from different agencies for 
the same entities coincide often but not always. By exception, the assessments by 
different agencies may differ 
significantly. Similarly, for samples of borrowers ranked by experts, the resulting rankings will 
depend more often than not on the selection of the experts and possibly 
also on the way the assessments by the single experts are combined to one ranking. 

So, as it proves difficult to compare the statistical rating model in question to `right' or most predictive 
ratings, why not comparing it to a perfect (or `prophetic') rating system?
This approach might seem strange at first glance, but it turns out to be viable -- even if
it delivers only an approximation to the natural override rate.

A perfect rating system obviously needs only two grades: Default and survival. 
We have to define a mapping from the rating grades that are assigned by the statistical rating model in question
to the two states of default and survival in order to be able to calculate which portion of the borrowers in the portfolio would have to be moved were the distribution of borrowers across the range of proposed grades to be 
transformed into a distribution according to a perfect rating system.
For this purpose, we assume that the perfect rating system for the comparison is `realised' on the rating scale of the 
rating model under consideration. This implies that there is a multitude of perfect rating systems that could be 
used as the target for the comparison. For each rating system that assigns defaulters and survivors to 
disjoint sets of grades is perfect.

To devise the mapping from the proposed ratings to the perfect ratings, inspection of the concept of investment and 
speculative grades employed by the major rating agencies proves helpful. Moody's define 
\emph{Investment grade} as the combination of grades Aaa, Aa, A, and Baa while they call 
\emph{Speculative grade} the combination of grades Ba, B, and Caa-C. Similarly, S\&P define 
\emph{Investment grade} as the combination of grades AAA, AA, A, and BBB and 
\emph{Speculative grade} as the combination of grades BB, B, and CCC-C. 

\begin{table}[t!p]
\caption{Average annual default rates as recorded by \citet[][Exhibit~35]{Moodys2011} 
and \citet[][Table~24]{S&P2011}.}
\label{tab:1}
\begin{center}
\begin{tabular}{|l||c|c|c|}
\hline
Agency & Investment grade & Speculative grade & Observation period\\ \hline \hline
Moody's & 0.095\% & 4.944\% & 1983-2010 \\ \hline
S\&P & 0.13\% & 4.36\%  & 1981-2010 \\ \hline
\end{tabular}
\end{center}
\end{table}
Table~\ref{tab:1}
presents long-run average annual default rates for investment and speculative grade borrowers
as observed by Moody's and S\&P. On the one hand, these default rates show that, from a
credit default perspective, Moody's and S\&P's investment grade and speculative grade definitions
are broadly equivalent. On the other hand, and more important for this paper, the observed
default rates support the description of investment grade borrowers as `very unlikely to default within
a year' and of speculative grade borrowers as `at significant risk of default within a year'.
One might even be tempted to talk about investment grade borrowers as `safe' borrowers 
and about speculative grade borrowers as `risky' borrowers. This can be seen as a reasonable 
approximation
of a perfect rating system with the two grades `defaulter' and `survivor' only.

Intuitively, a coarse classification of borrowers into safe and risky should be easier to achieve than the much
finer differentiation envisaged by the rating agencies and most internal rating systems of financial 
institutions. Indeed, the restrictions of overrides discussed in section~\ref{sec:mechanics}
might be rationalised by the consideration that it is primarily the safe vs.\ risky differentiation
that one has to get right.

We adopt this view for our approach to the identification of bounds for override rates with `natural'
error rates. Hence, we will define the natural error rate as the misclassification rate in a two-state 
coarse rating system that is derived from the statistical rating model in question by combining 
a suitable subset of grades to a `safe' super-grade and the complementary subset of grades to
a `risky' super-grade.

\subsection{Natural error rate}
\label{sec:errorRate}

We consider first how to determine the sets of scores or grades that define the two super-grades
safe and risky. Then we discuss how to calculate the corresponding misclassification rate that will be defined
as the natural error rate associated with the statistical rating model in question.

Actually, these two problems can be solved in a fully general context. Hence although in principle
we only need to investigate the cases of a rating variable with a finite range of values and of 
a continuous real-valued score variable, the major part of the following
discussion applies to any random variable $S$ with values in a general measurable space -- only minor
modifications of the notation might be necessary. 
In practice the measurable space will be a
multi-dimensional Euclidian space $\mathbb{R}^n$ when we are dealing with the risk factors
informing a rating model, an open real interval $I \subset \mathbb{R}$ when we are talking
about the score output of a rating model, or
a finite ordered set (without loss of generality $\{1, 2, \ldots, k\}$, with $k$ being
the number of performing rating grades) when the discussion is about
the ratings proposed by a rating model or the final ratings after overrides.

\textbf{Notation.} We assume that the conditional and unconditional distributions of $S$ have densities with
respect to a suitable measure. To make it clear when a statement or an equation 
is general we adopt the \emph{likelihood function} notation where a likelihood $\ell$ can stand both
for a \emph{probability density function} in a continuous or general context or for a
\emph{probability mass function} in a discrete context.

Speaking in technical terms, in this paper we study the joint distribution 
of a pair $(S, Z)$ of random variables. As mentioned before, the meaning of the variable $S$ could be
a vector of risk factors, a  \emph{rating model score}, or a
\emph{rating grade}, observed for a solvent borrower at a fixed date.
The variable $Z$ is the \emph{borrower's state of solvency} one
observation period (usually one year) after $S$ was observed. $Z$ takes on the two values 0 and 1.
The meaning of $Z=0$ is `borrower has remained solvent' (solvency or survival), 
$Z=1$ means `borrower has become insolvent' (default). 
We write $D$ for the event $\{Z=1\}$ and $N$ for the event $\{Z=0\}$.
Hence 
\begin{equation}\label{eq:D}
	D \cap N = \{Z=1\} \cap \{Z=0\} = \emptyset, \quad D \cup N = \text{whole space}.
\end{equation}
The marginal distribution of the state variable $Z$ is characterised by
the \emph{unconditional probability of default} $p$ which is defined as 
the probability of the default event $D$:
\begin{equation}\label{eq:PDunconditional}
	p = \mathrm{P}[D] = \mathrm{P}[Z =1] \in (0,1).
\end{equation}
The joint distribution of $(S, Z)$ then can be specified by
$p$ and the two distributions of $S$ conditional on the states of $Z$ (i.e.\
the events $D$ and $N$ respectively). Most important are the cases where
the conditional distributions are given by discrete
conditional probabilities $\ell_D(s) = \mathrm{P}[S = s\,|\,D]$ and
$\ell_N(s) = \mathrm{P}[S = s\,|\,N]$, $s = 1, \ldots, k$, or by Lebesgue densities
$\ell_D$ and $\ell_N$.  In the latter case, the probabilities of
$S$ taking a value in a set $M$ conditional on default and survival respectively can be written as
\begin{equation}\label{eq:densities}
	\mathrm{P}[S \in M\,|\,D] = \int_M \ell_D(s)\,ds \quad\text{and}\quad
	\mathrm{P}[S \in M\,|\,N] = \int_M \ell_N(s)\,ds.
\end{equation}

The joint distribution of the pair $(S, Z)$ of the score and the borrower's state one period later
can also be specified by the unconditional distribution $\mathrm{P}[S \in \cdot\,]$
of $S$ (i.e.\ the score or rating profile) and the PDs $\mathrm{P}[D\,|\,S] =
1-\mathrm{P}[N\,|\,S]$ \emph{conditional} on $S$. 
In the special cases we have mentioned before, by Bayes' rule we have the following representations
of the conditional PDs:
\begin{subequations}
\begin{itemize}
	\item $S$ is understood as rating grade, i.e.\ $S \in \{1, 2, \ldots, k\}$. Then
\begin{equation}\label{eq:pd_discrete}
		\mathrm{P}[D\,|\,S = s] \ =\ \frac{p\,\mathrm{P}[S = s\,|\,D]}{p\,\mathrm{P}[S = s\,|\,D] +
		(1-p)\,\mathrm{P}[S = s\,|\,N]}, \quad s \in \{1, 2, \ldots, k\}.
\end{equation}
	\item $S$ is a continuous score variable or a vector of continuous risk factors
	such that there are Lebesgue densities $\ell_N$ and $\ell_D$ 
	of the distributions conditional on default and survival as in \eqref{eq:densities}. Then	
\begin{equation}\label{eq:pd_continuous}
		\mathrm{P}[D\,|\,S = s]\ =\ \frac{p\,f_D(s)}{p\,f_D(s) + (1-p)\,f_N(s)}.
\end{equation}
\end{itemize}
With the likelihood notation, \eqref{eq:pd_discrete} and \eqref{eq:pd_continuous} can be expressed
in one equation:
\begin{equation}\label{eq:pd_lik}
		\mathrm{P}[D\,|\,S = s]\ =\ \frac{p\,\ell_D(s)}{p\,\ell_D(s) + (1-p)\,\ell_N(s)}.
\end{equation}
\end{subequations}
Thanks to \eqref{eq:pd_lik} the conditional PDs are determined
by the unconditional PD and the distributions of the score conditional on default and survival.
The conditional and the unconditional PDs are the information we need to determine the intended
split of the score or rating range into the two super-grades safe and risky. The split suggested
is based on the minimisation of the so-called \emph{expected misclassification cost}
\citep[see, e.g.,][]{hand2001simple}. 

\begin{proposition}\label{pr:loss}%
Assume that the expected cost of misclassifying a defaulting borrower as solvent is $c_D > 0$ and 
that the expected cost of misclassifying a solvent borrower as defaulting is $c_N > 0$. Define the 
events $D$ and $N$ as in \eqref{eq:D} and the unconditional PD $p$ as in \eqref{eq:PDunconditional}.
Assume that a borrower is classified as defaulting if 
and only if the borrower's score (or rating or vector of risk factors) $S$ takes on a value in a 
predefined set $A$. 
Then the expected cost $C$ of misclassification is given by
\begin{equation}\label{eq:criterion}
	C \ =\ c_D\,p\,\mathrm{P}[S \notin A\,|\,D] + c_N\,(1-p)\,\mathrm{P}[S \in A\,|\,N].
\end{equation}
The expected cost $C$ of misclassification is minimal if and only if the set $A$ is chosen as 
the following set $A^\ast$:
\begin{equation}\label{eq:decision}
	A^\ast \ = \ \left\{\mathrm{P}[D\,|\,S] > \frac{c_N}{c_N+c_D}\right\}.
\end{equation}
\end{proposition}
See, for instance, \citet[][Theorem~2]{nechvalmodified} for a proof of proposition~\ref{pr:loss}.
Note that \eqref{eq:decision} defines a risky super-grade in the sense that the average PD
conditional on the score or rating on the set $A^\ast$ is greater than the average PD 
conditional on the score or rating on the complement of the set $A^\ast$:
\begin{equation}\label{eq:comparison}
	\mathrm{E}\bigl[\mathrm{P}[D\,|\,S]\,|\,S\in A^\ast\bigr] \ >\  \frac{c_N}{c_N+c_D}
	\ \ge \ \mathrm{E}\bigl[\mathrm{P}[D\,|\,S]\,|\,S\notin A^\ast\bigr]. 
\end{equation}
How to choose the cost parameters $c_D$ and $c_N$ in \eqref{eq:decision}? \citet[][section~4]{hand2009measuring}
has commented that in general it is extremely difficult to determine values for the cost parameters. Even
as it obviously suffices to specify the ratio of the cost parameters, most of time it appears difficult to come to 
a meaningful assessment. 

Nonetheless, as noted by \citet[][section~4]{hand2009measuring}, chosing the cost parameters inversely proportional 
to the unconditional class probabilities is popular. In the setting of this paper, this means 
\begin{equation}\label{eq:cost}
	c_D \ =\ 1/p \qquad \text{and} \qquad c_N \ =\ 1/(1-p).
\end{equation}
In a context of risky lending, \eqref{eq:cost} is actually a quite reasonable choice. Assume that the
credit decision is about a loan with principal $M$ and that the expected loss rate in the case of default
is $0 < \lambda \le 1$. Then the effective (after cost of funding) interest 
charged for the loan should be at least a provision for the expected loss which equals $M\,\lambda\,p$. 
If we prudentially assume that the interest is paid
in advance then the lending institution's loss in the case of lending to a defaulter will be
\begin{subequations} 
\begin{equation}\label{eq:defaultLoss}
	M\,\lambda - M\,\lambda\,p \ =\ M\,\lambda\,(1-p).
\end{equation}
The loss in case of not lending to a customer who turns out not to be a defaulter will be (at least) $M\,\lambda\,p$.
Together with \eqref{eq:defaultLoss} this implies for the ratio of the misclassification cost parameters
\begin{equation}\label{eq:ratio}
	\frac{M\,\lambda\,(1-p)}{M\,\lambda\,p}\ =\ \frac{1/p}{1/(1-p)}\ = \ \frac{c_D}{c_N},
\end{equation}
\end{subequations}
with $c_D$ and $c_N$ as suggested in \eqref{eq:cost}. Even if the assumption on interest being paid in advance
clearly is often not satisfied in practice, for not too large PD $p$ the ratio of misclassification costs
will still be of a magnitude similar to the left-hand side of \eqref{eq:ratio}. We therefore adopt \eqref{eq:cost}
for the choice of the cost parameters in our application of proposition~\ref{pr:loss}.

Observe that with cost parameters as defined by \eqref{eq:cost}, by equations \eqref{eq:pd_lik} and \eqref{eq:decision}
the optimal set $A^\ast$ of grades or ratings for classifying borrowers as defaulting can be rewritten as
follows:
\begin{subequations} 
\begin{align}
A^\ast & = \left\{\frac{1}{1 + \bigl((1-p)/p\bigr)\,\bigl(\ell_N(S)/\ell_D(S)\bigr)} > \frac{1}{1+c_D/c_N}\right\} \notag\\
	& = \left\{\frac{\ell_D(S)}{\ell_N(S)} > \frac{1-p}{p}\,\frac{c_N}{c_D}\right\} \label{eq:lik_general}\\
	& = \left\{\frac{\ell_D(S)}{\ell_N(S)} > 1\right\}\label{eq:lik_special}.
\end{align}
\end{subequations} 
Comparison of \eqref{eq:lik_general} and \eqref{eq:lik_special} shows that the same optimal set $A^\ast$ 
for classifying borrowers as defaulting is found, no matter whether we start from a general
unconditional PD $p$ and cost parameters given by \eqref{eq:cost} or from $p = 50\%$ and equal cost
parameters $c_D = c_N$.

\begin{remark}
Assume that the variable $S$ from proposition~\ref{pr:loss} stands for a one-dimensional score or rating.
If the likelihood ratio $\frac{\ell_D(s)}{\ell_N(s)}$ is decreasing\footnote{%
See \cite{Tasche2008a} for a discussion of the connection between monotonicity
of the likelihood ratio and optimality of the model scores.} in $s$ (or, equivalently,
the PD conditional on the score $\mathrm{P}[D\,|\,S=s]$ is increasing in $s$) then 
the set $A^\ast$ from \eqref{eq:decision} indicating the classification of borrowers as defaulting can be
written as
\begin{equation}
\begin{split}
	A^\ast & = \{S \le s^\ast\}, \\
	s^\ast & = \sup\left\{s:\, \frac{\ell_D(s)}{\ell_N(s)} > 1\right\}.
\end{split}	
\end{equation}
In this case the minimum classification cost $C^\ast$ according to \eqref{eq:criterion} 
with cost parameters defined by \eqref{eq:cost} is equivalent to
the Kolmogorov-Smirnov statistic applied to the distribution functions of the score
or rating conditional on default and survival respectively:
\begin{equation}\label{eq:KS}
    C^\ast\ =\ 1 - \max_s \bigm|\mathrm{P}[S \le s\,|\,D] - \mathrm{P}[S \le s\,|\,N]\bigm|.
  \end{equation}
This is an interesting observation since it justifies, in economic terms, the use of the
Kolmogorov-Smirnov statistic and of the closely related accuracy ratio for the performance
measurement of rating systems. 
\end{remark} 
With \eqref{eq:lik_special} we have identified the decision rule that should be applied to
determine the two-state coarse rating system that was discussed in section~\ref{eq:benchmark}.
This rating system is an overlay to the statistical rating model whose output is
the variable $S$ we have considered in proposition~\ref{pr:loss} and its conclusions.
The misclassification rate of the overlaying rating system with the two super-grades `risky' and `safe' 
gives us a reasonable bound for the override rate of the statistical rating model $S$.

\begin{subequations}
\begin{proposition}\label{pr:natural}
Define the \emph{natural error rate} $\epsilon(S)$ of a statistical rating model with output $S$ 
as the misclassification rate associated with the decision rule \eqref{eq:lik_special}
for classifying borrowers as defaulting. Let $p$ denote the
unconditional probability of default. Then $\epsilon(S)$ is given by
\begin{equation}\label{eq:natural}
	\epsilon(S) \ =\ p\,\mathrm{P}[\ell_D(S) \le \ell_N(S)\,|\,D] + 
									(1-p)\,\mathrm{P}[\ell_D(S) > \ell_N(S)\,|\,N].
\end{equation}
\end{proposition}
Note that the natural error rate as defined by \eqref{eq:natural} is 
an \emph{actual error rate} in the terminology of \citet[][section~7.2]{Hand97}.
If the output of the statistical rating model is a rating proposal or 
a discrete score with values
in $\{1, \ldots, k\}$ then the likelihood functions in \eqref{eq:natural} are 
probability mass functions. Let
\begin{equation}\label{eq:J}
	J \ =\ \bigl\{j: \mathrm{P}[S=j\,|\,D] > \mathrm{P}[S=j\,|\,N]\bigr\}.
\end{equation}
Then \eqref{eq:natural} can be written as
\begin{equation}\label{eq:error.discrete}
	\epsilon(S) \ =\ p \sum_{j\notin J} \mathrm{P}[S=j\,|\,D] + (1-p) \sum_{j\in J} \mathrm{P}[S=j\,|\,N].
\end{equation}
If the output of the statistical rating model is a continuous score with values
in an open interval $I$ then the likelihood functions in \eqref{eq:natural} are 
probability density functions. \eqref{eq:natural} can then be represented as
\begin{equation}
	\epsilon(S) \ =\ p \int_{\{s\in I: \ell_D(s) \le \ell_N(s)\}} \ell_D(s)\,d s +
				(1-p) \int_{\{s\in I: \ell_D(s) > \ell_N(s)\}} \ell_N(s)\,d s.
\end{equation}
\end{subequations}
As mentioned before, the purpose of this paper is to demonstrate that the natural error 
rate of a statistical rating model as defined formally in proposition~\ref{pr:natural}
is a suitable bound for the rate of overrides that should be applied to the model's rating proposals.
In section~\ref{tas_sec_2}, we show that the natural error rate has the intuitive property that
its value increases when the discriminatory power of the statistical rating models declines.
This observation strengthens the case we have made in section~\ref{eq:benchmark} for
the adoption of the natural error rate as a bound for the rating override rate.

It is worthwhile noting, however, that a regrouping of borrowers according to the
decision rule \eqref{eq:lik_special} underlying the coarse approximate rating system would
imply moving a much larger portion of solvent borrowers from the `risky'
grade to the `safe' grade than of defaulting borrowers from safe to risky.
This observation might seem counterintuitive at first glance. 
Nonetheless, in practice the movement
from risky to safe would be quite restrictived as only borrowers that
have been very clearly misclassified should be moved. This consideration
indicates that the `natural error rate' indeed is an upper bound for the
rating override rate since most of the misclassified borrowers cannot be
identified with reasonable certainty.


\section{Override rate and discriminatory power}
\label{tas_sec_2}

Define the conditional distribution functions $F_D$ and $F_N$ of the score or rating variable $S$
described in section~\ref{sec:errorRate} by
\begin{equation}
	F_D(s) \ =\ \mathrm{P}[S \le s\,|\,D] \qquad \text{and}\qquad
	F_N(s) \ =\ \mathrm{P}[S \le s\,|\,N],
\end{equation}
and denote by $S_D$ and $S_N$ independent random variables that are distributed 
according to $F_D$ and $F_N$ respectively.

For quantifying discriminatory power, we apply the notion of \emph{accuracy ratio (AR)} as specified in \citet[][eq.~(3.28b)]{Tasche2009a}:
\begin{equation}\label{eq:AR}
\begin{split}
	\mathrm{AR} &  = 2\,\mathrm{P}[S_D < S_N] + \mathrm{P}[S_D = S_N] - 1 \\
	& = \mathrm{P}[S_D < S_N] - \mathrm{P}[S_D > S_N]\\
	&= \int \mathrm{P}[S < s\,|\,D]\,d F_N(s) - \int \mathrm{P}[S < s\,|\,N]\,d F_D(s),
\end{split}
\end{equation}
See \citet[][section~2]{hand2001simple} for a discussion of why this definition of accuracy ratio (or
the related definition of the area under the ROC curve) is more expedient than the also common definition
in geometric terms. Definition \eqref{eq:AR} of AR takes an `ex post' perspective by assuming the 
obligors' states $D$ or $N$ one year after having been scored are known and hence 
can be used for estimating the conditional (on default and survival respectively) 
score distributions $F_D$  and $F_N$.

\subsection{The binormal case}

The ex-post perspective on discriminatory power is appropriate 
when we consider the important binormal special case where both
conditional score distributions are normal distributions with different mean values but equal
variances. This special case, in particular, motivates the choice of the inverse logit function
for modelling PD curves \citep[see, e.g.,][section 6.1]{Cramer2003} 
since in practice it is often reasonable to assume that the two 
conditional score distributions are approximately normal.

\begin{subequations}
\begin{proposition}\label{pr:normal}
Assume that the conditional distributions of the score variable $S$ are given by
$S_D \sim \mathcal{N}(\mu_D, \sigma)$ and $S_N \sim \mathcal{N}(\mu_N, \sigma)$
with $\mu_D \le \mu_N$. Let $p$ denote the unconditional probability of default as
given by \eqref{eq:PDunconditional}. 
Then the accuracy ratio $AR$ associated with $S$ according to \eqref{eq:AR} can be calculated as\footnote{%
$\Phi$ denotes the standard normal distribution function.}
\begin{equation}\label{eq:AR.normal}
	AR \ =\ 2\,\Phi\biggl(\frac{\mu_N - \mu_D}{\sqrt{2}\,\sigma}\biggr) - 1.
\end{equation}
The natural error rate $\epsilon(S)$ of $S$ according to proposition~\ref{pr:natural} is given
by
\begin{equation}\label{eq:error.normal}
	\epsilon(S) \ = \ \Phi\biggl(\frac{\mu_D - \mu_N}{2\,\sigma}\biggr).
\end{equation}
With the cost parameters defined by
\eqref{eq:cost}, the average conditional PDs on the set of scores indicating
default and on the set of scores indicating survival respectively (as in eq.~\eqref{eq:comparison})
are given by
\begin{equation}\label{eq:PDaverage}
\begin{split}
\mathrm{E}\bigl[\mathrm{P}[D\,|\,S]\,|\,S\in A^\ast\bigr] & = 
	\frac{p\,(1-\epsilon(S))}{p\,(1-\epsilon(S)) +
			(1-p)\,\epsilon(S)},\\
\mathrm{E}\bigl[\mathrm{P}[D\,|\,S]\,|\,S\notin A^\ast\bigr] & = 
	\frac{p\,\epsilon(S)}{p\,\epsilon(S) +
			(1-p)\,(1-\epsilon(S))}.
\end{split}	
\end{equation}
\end{proposition}
\end{subequations}
\textbf{Proof.} \eqref{eq:AR.normal} follows from \citet[][eq.~(3.14) and proposition~3.15]{Tasche2009a}.
For \eqref{eq:error.normal} and \eqref{eq:PDaverage}, observe that 
in the binormal case with equal variances we have 
\begin{equation*}
	S\notin A^\ast \quad \iff \quad \ell_D(S) \le \ell_N(S) \quad \iff \quad S \ge \frac{\mu_D+\mu_N}2.
\end{equation*}
From this observation, we deduce that
\begin{align*}
	\mathrm{P}[\ell_D(S) \le \ell_N(S)\,|\,D] & = 1-\Phi\biggl(\frac{\mu_N - \mu_D}{2\,\sigma}\biggr)\\
	\intertext{and}
	\mathrm{P}[\ell_D(S) > \ell_N(S)\,|\,N] & = \Phi\biggl(\frac{\mu_D - \mu_N}{2\,\sigma}\biggr).
\end{align*}
These equations imply \eqref{eq:error.normal}. 
With regard to \eqref{eq:PDaverage}, we apply the definition of conditional expectation to arrive
at the following computation\footnote{We define the indicator function
$\mathbf{1}_M$ of a set $M$ by $\mathbf{1}_M(m) = \begin{cases}
	1, & m \in M,\\
	0, & m \notin M.
\end{cases}$}:
\begin{align*}
	\mathrm{E}\bigl[\mathrm{P}[D\,|\,S]\,|\,S\in A^\ast\bigr] & = 
		\frac{\mathrm{P}\bigl[\mathrm{P}[D\,|\,S]\,\mathbf{1}_{\{S\in A^\ast\}}\bigr]}{\mathrm{P}[S\in A^\ast]}\\
	& = \frac{\mathrm{P}[D\cap \{S < \frac{\mu_D+\mu_N}2\}]}{\mathrm{P}[S < \frac{\mu_D+\mu_N}2]}\\
	& = \frac{\mathrm{P}[S < \frac{\mu_D+\mu_N}2\,|\,D]\,\mathrm{P}[D]}{\mathrm{P}[S < \frac{\mu_D+\mu_N}2\,|\,D]\,\mathrm{P}[D] + \mathrm{P}[S > \frac{\mu_D+\mu_N}2\,|\,N]\,\mathrm{P}[N]}\\
	& = \frac{p\,\Phi\bigl(\frac{\mu_N - \mu_D}{2\,\sigma}\bigr)}{p\,\Phi\bigl(\frac{\mu_N - \mu_D}{2\,\sigma}\bigr) +
			(1-p)\,\Phi\bigl(\frac{\mu_D - \mu_N}{2\,\sigma}\bigr)}.
\end{align*}
Since $\Phi\bigl(\frac{\mu_N - \mu_D}{2\,\sigma}\bigr) = 1-\epsilon(S)$ we obtain the first
equation in \eqref{eq:PDaverage}. The second equation follows in the same way.
\hfill $\Box$

In proposition~\ref{pr:normal}, it is a general property of the accuracy ratio that it does not depend 
on the unconditional PD. The observation that the natural error rate does not depend on the unconditional
PD might be surprising at first glance. This observation, however, is primarily a consequence of the fact
that both conditional score distributions are members of the same location-scale distribution
family, different only by the different location parameters. In addition, symmetry of the
location zero member of the family is required to imply the property that the error does not depend on the
unconditional PD. 

\begin{figure}[t!p]
\caption{Natural error rate as function of the discriminatory power (accuracy ratio) 
for the binormal case described in corollary~\ref{co:normal} and 
for the discrete rating model described in example~\ref{ex:discrete}.}
\label{fig:2}
\begin{center}
\ifpdf
	\includegraphics[width=15cm]{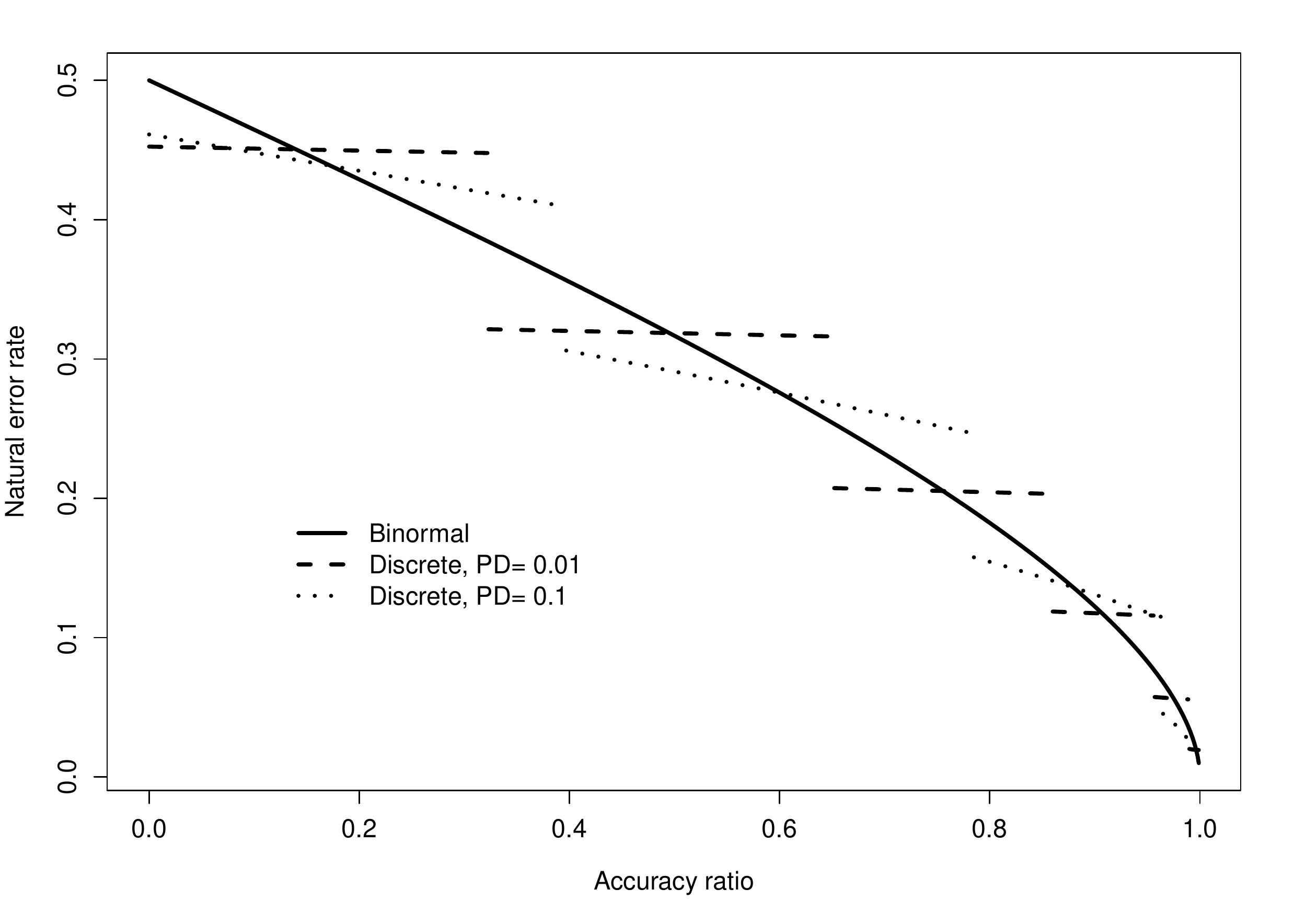}
\fi
\end{center}
\end{figure}

Combining \eqref{eq:AR.normal} and \eqref{eq:error.normal} provides a simple relationship
between the discriminatory power as measured by the accuracy ratio and the natural error
rate in the binormal case with equal variances. Eq.~\eqref{eq:main} shows, in particular, that
the natural rating error indeed decreases when the discriminatory power of the statistical
rating model increases.

\begin{corollary}\label{co:normal}
Assume that the conditional distributions of the score variable $S$ are given by
$S_D \sim \mathcal{N}(\mu_D, \sigma)$ and $S_N \sim \mathcal{N}(\mu_N, \sigma)$
with $\mu_D \le \mu_N$. Then the accuracy ratio $AR$ and the natural error rate $\epsilon(S)$
associated with $S$ are related by the following equation:
\begin{equation}\label{eq:main}
	\epsilon(S) \ = \ \Phi\left(- \,\frac 1{\sqrt{2}}\,\Phi^{-1}\Big(\frac{AR+1}2\Big)\right).
\end{equation}
\end{corollary}
See figure~\ref{fig:2} for a graphical representation of relation \eqref{eq:main}.

\subsection{The discrete case}

For practical applications, often one cannot assume that the conditional score distributions
are normally distributed. Moreover, as described in section~\ref{sec:mechanics}, it is actually
not the score output of a statistical rating model that may be overridden but the proposed ratings
derived from the scores. Typically, thresholds for override rates for a statistical rating model 
are established when the development of the model is being finished and the documentation of the
related rating system is compiled. At this stage, it is unlikely that reliable estimates 
of the score distribution conditional on the borrower's solvency state or of the proposed rating distributions
conditional on the borrower's solvency state are available. Therefore, we also consider the case where
the unconditional distribution of the proposed ratings (rating profile) and the associated 
PD curve (i.e.\ the estimates of the PDs conditional on the proposed ratings) are all that is known.

Nonetheless, with this information it is still possible to compute both the natural error rate and
the accuracy ratio for the statistical rating model in question. Key to these computations is the
following observation that eq.~\eqref{eq:pd_discrete} can be `inverted' such that the distributions
of the proposed ratings conditional on the borrower's solvency state can be derived from the
unconditional rating profile and the PD curve:
\begin{equation}\label{eq:inverse}
\begin{split}
	p  &\ =\ \mathrm{E}\bigl[\mathrm{P}[D\,|\,S]\bigr]\ =\ 
	\sum_{s=1}^k \mathrm{P}[D\,|\,S = s]\,\mathrm{P}[S = s],\\
\mathrm{P}[S = s\,|\,D]  &\ =\ \mathrm{P}[D\,|\,S = s]\,\mathrm{P}[S = s] / p, \quad s = 1, \ldots, k,\\
\mathrm{P}[S = s\,|\,N]  &\ =\ \bigl(1-\mathrm{P}[D\,|\,S = s]\bigr)\,\mathrm{P}[S = s] / (1-p),
\quad s = 1, \ldots, k.
\end{split}
\end{equation}
By \eqref{eq:inverse} and \eqref{eq:error.discrete}, the natural error rate can be calculated from
the profile $s \mapsto \mathrm{P}[S = s]$ of the proposed ratings and the PD curve 
$s \mapsto \mathrm{P}[D\,|\,S = s]$.
By combining \eqref{eq:AR} and \eqref{eq:inverse}, also the accuracy ratio
of the ratings proposed by the statistical model can be determined from the rating profile
and the PD curve:
\begin{equation}\label{eq:AR.discrete}
\begin{split}
	AR & = \frac 1{p\,(1-p)} \Big(2 \sum_{s=1}^k \bigl(1-\mathrm{P}[D\,|\,S=s]\bigr)\,\mathrm{P}[S=s]\sum_{t=1}^{s-1} 
	\mathrm{P}[D\,|\,S=t]\,\mathrm{P}[S=t]\\
	&	\quad +
	 \sum_{s=1}^k \mathrm{P}[D\,|\,S=s]\,\bigl(1-\mathrm{P}[D\,|\,S=s]\bigr)\,\mathrm{P}[S=s]^2\Big) \ - 1.
\end{split}	 
\end{equation}
Calculation of the accuracy ratio by means of the combination of \eqref{eq:inverse} and \eqref{eq:AR.discrete}
may be interpreted as predicting the accuracy ratio `ex ante' since it can be done on the current portfolio
as soon as estimates of the PDs conditional on the rating grades (PD curve) are available.

\begin{figure}[t!p]
\caption{PD curves for the rating model described in example~\ref{ex:discrete}. The curves
have been inferred from the unconditional rating distribution shown in figure~\ref{fig:1} under
the assumption that the unconditional PD is 5\% and the accuracy ratio associated with the rating 
model is either low at 25\% or high at 75\%.}
\label{fig:3}
\begin{center}
\ifpdf
	\includegraphics[width=15cm]{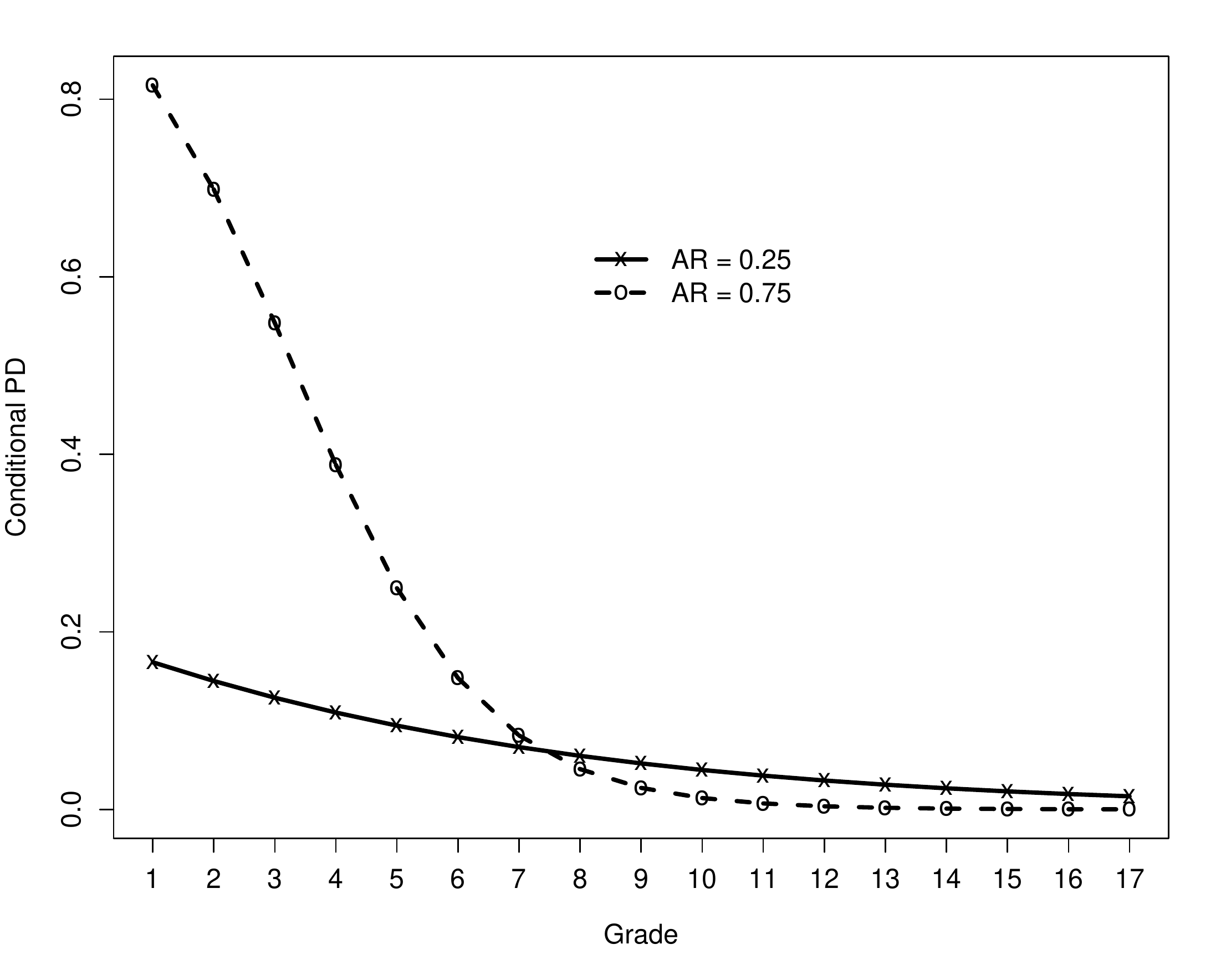}
\fi
\end{center}
\end{figure}
There is no obvious general example of a discrete rating model comparable to the binormal model 
from corollary~\ref{co:normal} that would allow the direct study
of the connection between discriminatory power and natural error rate.
We therefore explore this connection in the discrete case by means of a specific example.
In order to make the example differ significantly from the normal assumption discussed before 
it has been chosen in a way as to create a certain degree of unsymmetry and overdispersion.

\begin{example}\label{ex:discrete}
The unconditional distribution of the ratings on a discrete scale from 1 to 17 is given by a correlated binomial
distribution. That is if $S$ stands for a borrower's rating grade, $S$ can be written
as $S = X + 1$ where the distribution of $X$ is specified by\footnote{%
$\varphi$ denotes the standard normal density function: $\varphi(t) = \frac{e^{- t^2/2}}{\sqrt{2\,\pi}}$.}
\begin{subequations}
\begin{equation}\label{eq:corrbinomial}
\begin{split}
	\mathrm{P}[X \le x] & = \int\limits_{-\infty}^\infty \varphi(y)\,\sum_{i=0}^x
\left(\begin{smallmatrix}
  k\\ i
\end{smallmatrix}\right)
\,G(\lambda,\varrho, y)^i\,(1-G(\lambda,\varrho, y))^{k-i}\,d\,y, \quad x = 0, \ldots k,\\
G(\lambda,\varrho, y) & =
   \Phi\Big(\frac{\Phi^{-1}(\lambda)-\sqrt{\varrho}\,y}{\sqrt{1-\varrho}}\Big).
\end{split}
\end{equation}
For the purpose of this example we chose the following values for the parameters
$k$ (number of grades minus one), $\lambda$ (determines mean of the distribution)
and $\varrho$ (drives overdispersion compared to binomial distribution) in \eqref{eq:corrbinomial}:
\begin{equation}
	k = \mathrm{16}, \quad \lambda = 0.55, \quad \varrho = 0.1.
\end{equation}
\end{subequations} 
We assume that the PD conditional on a rating grade $S=s$ is appropriately discribed by the inverse logit
function:
\begin{equation}
	\mathrm{P}[D\,|\,S=s]\ =\ \frac{1}{1+e^{a+b\,s}}, \quad s=1, \ldots, 17.
\end{equation}
The parameters $a$ and $b$ are determined by \emph{quasi-moment matching} \citep[][section~5.2]{Tasche2009a}.
This approach works by equating  the right-hand side of the first equation of \eqref{eq:inverse} and 
the right-hand side of \eqref{eq:AR.discrete} to predefined values of $PD$ and $AR$ respectively and solving
numerically for $a$ and $b$:
\begin{align}
	PD & = \sum_{s=1}^{\mathrm{17}} \mathrm{P}[S=s]\,\mathrm{P}[D\,|\,S=s]  =  \sum_{s=1}^{\mathrm{17}} 
	\frac{\mathrm{P}[X=s-1]}{1+e^{a+b\,s}},\notag\\
	AR &\ =\ \frac 1{PD\,(1-PD)} \Big(2 \sum_{s=1}^{\mathrm{17}} \frac{e^{a+b\,s}}{1+e^{a+b\,s}}\,\mathrm{P}[X=s-1]\sum_{t=1}^{s-1} 
	\frac{\mathrm{P}[X=t-1]}{1+e^{a+b\,t}}\label{eq:quasi}\\
	&	\quad +
	 \sum_{s=1}^{\mathrm{17}} \frac{e^{a+b\,s}}{\bigl(1+e^{a+b\,s}\bigr)^2}\,\mathrm{P}[X=s-1]^2\Big) \ - 1.
	 \notag
\end{align}
\end{example}
\begin{table}[t!p]
\caption{Natural error rate as function of the discriminatory power (accuracy ratio) 
for the binormal case described in corollary~\ref{co:normal} and 
for the discrete rating model described in example~\ref{ex:discrete}. See
figure~\ref{fig:2} for a graphical representation.}
\label{tab:2}
\begin{center}
\begin{tabular}{|c||c|c|c|}
\hline
AR & \multicolumn{3}{|c|}{Natural error rate}\\ \hline
 & Binormal & Discrete, PD=1\% & Discrete, PD=10\% \\ \hline \hline
0.0 & 0.500 & 0.452 & 0.461 \\ \hline
0.1 & 0.465 & 0.451 & 0.448 \\ \hline
0.2 & 0.429 & 0.450 & 0.435 \\ \hline
0.3 & 0.393 & 0.448 & 0.422 \\ \hline
0.4 & 0.355 & 0.320 & 0.306 \\ \hline
0.5 & 0.317 & 0.319 & 0.291 \\ \hline
0.6 & 0.276 & 0.317 & 0.276 \\ \hline
0.7 & 0.232 & 0.206 & 0.260 \\ \hline
0.8 & 0.182 & 0.204 & 0.154 \\ \hline
0.9 & 0.122 & 0.118 & 0.131 \\ \hline
\end{tabular}
\end{center}
\end{table}

Figure~\ref{fig:3} illustrates the results of quasi-moment matching according to \eqref{eq:quasi}.
In particular, it becomes clear that the slope of the PD curve is primarily controlled by the discriminatory
power of the rating model as expressed by its accuracy ratio. The two curves from figure~\ref{fig:3} have
been used together with the unconditional rating distribution as specified by \eqref{eq:corrbinomial}
to calculate (by means of \eqref{eq:inverse}) the conditional rating distributions that are shown in figure~\ref{fig:1}.
Once the conditional rating distributions have been determined, the natural error rates (and hence
bounds for the overrides) may be computed by means of equations \eqref{eq:J} and \eqref{eq:error.discrete}.

\subsection{Comparing the binormal and the discrete results}

The same approach that was used for the calculation of the conditional rating distributions in figure~\ref{fig:1}
has also been applied in order to calculate the discrete error rate curves for figure~\ref{fig:2},
albeit with different parameters. For figure~\ref{fig:2}, the calculations were done for a small unconditional
PD (1\%) and
a large unconditional PD (10\%). Each of the two PDs was combined with the whole range $[0, 1)$ of potential ARs to 
first determine conditional PD curves by means of \eqref{eq:quasi} and then natural error rates by means of 
\eqref{eq:J} and \eqref{eq:error.discrete}. The discontinuities in the `discrete' curves are owed to 
changes in the index set $J$ from \eqref{eq:J} which cause jumps in the values of the error rates. Despite
the jumps, the two discrete curves are remarkably close to the binormal curve from corollary~\ref{co:normal}.
This observation is confirmed by table~\ref{tab:2} that provides the numerical values for some of the points on the 
three curves in figure~\ref{fig:2}. It appears therefore worthwhile to consider that the `normal' values
calculated with \eqref{eq:main} are taken to indicate the link between discriminatory power
and override even for non-normal cases.
\begin{table}[t!p]
\caption{Average `safe' and `risky' PDs as function of the discriminatory power (AR)
for the binormal case \eqref{eq:PDaverage} and
example~\ref{ex:discrete}. The assumed unconditional PD is 1\%.}
\label{tab:3}
\begin{center}
\begin{tabular}{|c||c|c|c|c|}
\hline
AR (\%) & \multicolumn{2}{|c|}{Binormal} & \multicolumn{2}{|c|}{Discrete}\\ \hline
 & Safe PD (\%) & Risky PD (\%) & Safe PD (\%) & Risky PD (\%) \\ \hline \hline
0 & 1 & 1 & 1 & 1 \\ \hline
10 & 0.87 & 1.15 & 0.87 & 1.16 \\ \hline
20 & 0.75 & 1.33 & 0.74 & 1.32 \\ \hline
30 & 0.65 & 1.54 & 0.61 & 1.47 \\ \hline
40 & 0.55 & 1.8 & 0.57 & 1.9 \\ \hline
50 & 0.47 & 2.13 & 0.45 & 2.15 \\ \hline
60 & 0.38 & 2.58 & 0.34 & 2.39 \\ \hline
70 & 0.3 & 3.24 & 0.32 & 3.54 \\ \hline
80 & 0.22 & 4.33 & 0.19 & 4.02 \\ \hline
90 & 0.14 & 6.75 & 0.13 & 7.06 \\ \hline
\end{tabular}
\end{center}
\end{table}

In section~\ref{eq:benchmark}, the approach to override rate bounds via the misclassification rate
of a coarse two-state rating system was motivated by an inspection of the investment and speculative grades
concept used by the major rating agencies. By table~\ref{tab:1}, we had noted that investment grade
may be interpreted as `safe' and speculative grade may be regarded as `risky'. Table~\ref{tab:3}
presents `risky' and `safe' grade PDs in the sense of equation~\eqref{eq:comparison}
for the normal and discrete examples considered in this section.

Note the similarity between the average PDs from the 80\% accuracy ratio row of table~\ref{tab:3} and
the observed investment and speculative grade default rates from table~\ref{tab:1}. Moody's
report an overall average annual default rate of 1.8\% \citep[][Exhibit~35]{Moodys2011} and
an average one-year accuracy ratio of c.~85\% \citep[][Exhibit~15]{Moodys2011}. S\&P report 
an overall average annual default rate of 1.6\% \citep[][Table~24]{S&P2011} and
an average one-year accuracy ratio of c.~84\% \citep[][Table~2]{S&P2011}. Hence the observed similarity
between average PDs on the `risky' and `safe' super-grades and the investment and speculative
grade default rates recorded by the agencies might not be an incident. Rather one might guess
that there is an expected cost concept like the one presented in proposition~\ref{pr:loss}
behind the agencies' investment versus speculative classification.

\section{Monitoring rating overrides}
\label{tas_sec_monitoring}

In order to put the considerations from the previous sections into context, in this section we  
describe an illustrative framework for the monitoring of rating overrides. In particular, we make
suggestions for the
role that the natural error rate defined in proposition~\ref{pr:natural} could play as a bound (limit) 
for the override rate in such a framework. 

Let us start by recalling the observations on
statistical rating models and rating overrides that have lead us to suggesting the natural error rate
as possible limit for rating override rates.
\begin{itemize}
	\item No statistical rating model can predict credit defaults with certainty. In particular if the
	financial consequences of wrong rating decisions can be serious, therefore, rating proposals by
	models should be scrutinised by credit experts and be overridden if necessary.
	\item However, also the credit experts can be wrong with their decisions. Hence, there is a need
	to monitor the performance of the overrides, too.
	\item On a single case basis, it is not possible to decide whether a rating decision was right or wrong.
	Even an override of a CCC rating proposal to a AAA rating, followed immediately by the borrower's default,
	could in principle be just an occurence of bad luck.
	\item Nonetheless, on the basis of the expected cost of misclassification, it is possible to 
	identify sets of 'safe' and 'risky' rating grades, thus overlaying a 'coarse' rating system onto 
	the original system with a finer rating scale. The safe and risky super-grades represent a 'will survive'
	and a 'will default' prediction respectively.
	\item The expected error rate of the 'coarse' rating system is called \emph{natural error rate}. 
	If the application of overrides is restricted to fundamental errors (i.e.\ safe borrowers are 
	classified as risky or risk borrowers are classified as safe) the natural error rate is also a 
	\emph{natural override rate}. 
	\item On portfolios with sufficient numbers of default observations, the natural error rate can 
	be directly estimated (ex post). On portfolios with low numbers of default observations, the natural error
	rate can still be inferred from the rating profile and the PD curve of the rating model (ex ante).
	\item An observed override rate higher than the natural error rate could be a consequence
	of unnecessary and potentially erroneous overrides or of an overestimation of the discriminatory
	power of the rating model. In either case, the rating process would have been found to operate
	sub-optimally.
	\item An observed override rate lower than the natural error rate could be a consequence
	of overlooked rating errors or of an underestimation of the discriminatory
	power of the rating model. No clear immediate conclusion on the well- or malfunction of the rating
	process would be possible.
\end{itemize}

Based on these observations and with a view on affirming correct calibration of the rating model, 
an overriding monitoring framework could include the following elements:

\textbf{At the beginning of the observation period (typically one year)}
\begin{itemize}
	\item Infer the natural error
	rate from the rating profile and the PD curve, by making use of \eqref{eq:error.discrete} and \eqref{eq:inverse},
	or from the accuracy ratio, by \eqref{eq:main}.
	\item Discourage credit experts from minor rating override. Only fundamental rating errors 
	should be amended by overrides. 
\end{itemize}
\textbf{At the end of the observation period}
\begin{itemize}
	\item If there is a sufficient number of default observations, the discriminatory power of the rating system
	pre and post overrides should be computed. If the discriminatory power post overrides is significantly
	lower than the power pre overrides an update of the override governance might be necessary.
	\item If the observed override rate exceeds the natural error rate, the consistency of 
	the PD curve with the observed discriminatory power should be checked (if there is a sufficient
	number of defaults) by comparing it to the ex ante accuracy ratio \eqref{eq:AR.discrete} 
	because the slope of the PD curve might be too large. If the number of observed defaults is not sufficient
	for the ex post estimation of discriminatory power it should be confirmed that the overrides were justified
	(e.g.\ by interviews with the credit experts). If this is confirmed the PD curve must be amended
	as to reflect the lower than expected discriminatory power.
	\item If the observed override rate is lower than the natural error rate it should be confirmed (e.g.\ by
	internal audit) that the due procedure is followed with regard to overrides. If this is the case
	amendment of the PD curve as to reflect the higher than expected discriminatory power could be
	considered.
	\item An imbalance of upward and downward rating overrides with more upward overrides might
	be an indication of PD overestimation. If further analyis finds (e.g. based on statistical tests
	pre and post overrides
	if there is a sufficient number of default observations) that there is no overestimation 
	an update of the override governance might be necessary.
	\item An imbalance of upward and downward rating overrides with more downward overrides might
	be an indication of PD underestimation. If further analyis finds  that there is no underestimation 
	an update of the override governance might be necessary.
\end{itemize}


\section{Conclusions}
\label{tas_sec_4}

The first part of this paper presents a suggestion of how to determine bounds for the rates of overrides
of rating proposals determined by statistical rating models. In practice, often the guidance for such overrides 
tends to restrict them to essential cases where the model's assessments of borrowers as
risky or safe might not be correct. Motivated by this observation, the suggested bound for
the override rate is the misclassification rate of a coarse rating system with the two grades
`risky' and `safe'. This coarse rating system is fed by the output of the rating model
in question and combines its rating grades to two super-grades, thereby minimising
the expected cost of misclassification. The rate of misclassifications with the coarse rating
system is called `natural error rate'. It is argued that the natural error rate is an appropriate
bound for the override rate of the statistical rating model in question.

In the second part of the paper, methods for determining the natural error rate 
and its use in an illustrative framework for the monitoring of overrides 
are discussed. It 
turns out that there is a particularly simple formula (eq.~\eqref{eq:main}) for the natural error rate if
the rating distributions conditional on the borrower's solvency state are both normal with equal variance
(binormal case). Formula \eqref{eq:main} for the natural error rate suggests that the natural error
rate decreases when the disciminatory power of the rating model increases -- as it should intuitively be
expected. We compare the results from the `binormal' formula to results for the natural error rate from
a more realistic discrete-valued example of a rating model. The comparison shows the results by both approaches
to be very close. This observation indicates that the binormal formula might be used in general 
as a rule of thumb
for deriving a bound for the override rate from the discriminatory power of the rating model. 

As demonstrated in the paper, the natural error rate of a rating model can be calculated both ex post, based on 
the discriminatory power realised during a previous observation period, and ex ante at the
beginning of the observation period. In the latter case one applies Bayes' rule to determine the 
discriminatory power implied by the combination of rating distribution and PD curve associated with
the rating model. Both the ex post and the ex ante methods to the computation of 
the natural error rate as a bound for the override rate
are promising tools for validating and monitoring the performance of statistical rating models.


\end{document}